\documentclass[nonatbib]{article}
\PassOptionsToPackage{numbers, compress}{natbib}

\usepackage[preprint]{neurips_2020}

\usepackage[utf8]{inputenc} 
\usepackage[T1]{fontenc}    
\usepackage{hyperref}       
\usepackage{url}            
\usepackage{booktabs}       
\usepackage{amsfonts}       
\usepackage{nicefrac}       
\usepackage{microtype}      
\usepackage{enumitem}
\usepackage{amsmath}
\usepackage{algorithmic}
\usepackage{algorithm}

\usepackage{amsmath}
\usepackage{amsthm}
\usepackage{amssymb}
\usepackage{algorithm}
\usepackage{subcaption}
\usepackage{color}
\usepackage[english]{babel}
\usepackage{graphicx}
\usepackage{grffile}
\usepackage{natbib}
\usepackage{wrapfig,epsfig}
\usepackage{epstopdf}
\usepackage{url}
\usepackage{color}
\usepackage{epstopdf}
\usepackage{multirow}
\usepackage[T1]{fontenc}
\usepackage{bbm}
\usepackage{comment}
\usepackage{dsfont}
\usepackage{makecell}
\usepackage{enumitem}

\DeclareMathOperator*{\argmax}{argmax}
\newtheorem{defn}{Definition}

\title{Climbing the WOL: Training for Cheaper Inference}

\author{
  Zichang Liu\thanks{First two authors have equal contribution} \\
  Rice University\\
  \texttt{zichangliu@rice.edu} \\
  \And
  Zhaozhuo Xu\textsuperscript{*} \\
  Rice University \\
  \texttt{zx22@rice.edu} \\
  \AND
  Alan Ji \\
  Rice University \\
  \texttt{abj3@rice.edu} \\
  \And
  Jonathan Li \\
  Stanford University \\
  \texttt{jlli@stanford.edu} \\
  \And
  Beidi Chen \\
  Rice University \\
  \texttt{Beidi.chen@rice.edu} \\
  \And
  Anshumali Shrivastava \\
  Rice University \\
  \texttt{anshumali@rice.edu} \\
}

\begin{document}
\maketitle

\begin{abstract}


Efficient inference for wide output layers (WOLs) is an essential yet challenging task in large scale machine learning. Most approaches reduce this problem to approximate maximum inner product search (MIPS), which relies heavily on the observation that for a given model, ground truth labels correspond to logits of highest value during full model inference. However, such an assumption is restrictive in practice. In this paper, we argue that approximate MIPS subroutines, despite having sub-linear computation time, are sub-optimal because they are tailored for retrieving large inner products with high recall instead of retrieving the correct labels. With WOL, the labels often have moderate inner products, which makes approximate MIPS more challenging. We propose an alternative problem formulation, called Label Superior Sampling (LSS), where the objective is to tailor the system to ensure retrieval of the correct label. Accordingly, we propose a novel learned hash approach, which is significantly more efficient and sufficient for high inference accuracy than MIPS baselines. Our extensive evaluation indicates that LSS can match or even outperform full inference accuracy with around $ \bf{5} \times$ speed up and  \bf{87} $\% $ energy reduction.

\end{abstract}

\section{Introduction}
\label{sec:intro}




In recent years, neural networks with wide output layers have obtained promising results in various applications such as recommendation systems~\cite{xue2017deep,Bhatia16,fan2019mobius} and language modeling~\cite{bengio2003neural,mikolov2010recurrent,mikolov2013efficient}. One of the significant challenges of deploying such models lies in the massive computation cost of giant matrix multiplications in wide output layers (WOLs), which can easily contain millions of neurons~\cite{Bhatia16}. To tackle this problem, many existing methods focus on one principle direction: reducing computation by approximating the full WOL output with a few representative logits. 
A common formulation for this direction in literature is to pose the problem as a Maximum Inner Product Search (MIPS)~\cite{shrivastava2014asymmetric} problem and exploit MIPS approximation algorithms~\cite{shrivastava2014improved,guo2016quantization,zhang2018navigating}. Specifically, each input embedding from the previous hidden layer serves as a query, and neurons from the output layer are treated as data. The goal is to find the top-k neurons that have the maximum inner product with the query in sub-linear time. The model then performs predictions based on the logits computed by the query and only the selected neurons. Current approximate MIPS algorithms focus on indexing the data via different data structures. In this way, search computation is largely reduced. As expected, there is a trade-off between search efficiency and accuracy.


{\bf Shortcomings of approximate MIPS formalism:} An approximate MIPS is a natural formulation for WOL inference since the inference phase of neural networks (NNs) treats logits (a monotonic function of the inner product between data embeddings and label neurons) as the score of the label~\cite{shrivastava2014asymmetric}. As a result, the topmost inner product is indeed the correct label, justifying the need to approximate the MIPS subroutine. However, the approximation brings a new source of trade-offs. The hardness of MIPS or any near-neighbor solutions depends on two things: 1) the value of the inner product of the correct class we want to search and 2) the gap between the best inner product values and the values of other inner products in consideration. Often with a large number of classes, the inner product values, even for the correct class, are significantly smaller. At the same time, there are many other classes with roughly the same inner products. As a result, it is unreasonable to expect any approximate MIPS sub-routine to efficiently retrieve the correct labels accurately. The final prediction of the network is very sensitive to this retrieval accuracy.

{\bf Better retrieval can even beat full softmax:} We all know that NN models cannot generalize perfectly to testing data in practice. Specifically, the correct class might not have the maximum prediction score. Consider some statistics from Delicious-200K dataset\cite{deli200k}: the average rank of label neurons in inner products is only 498.14 out of 205443 during the inference phase of a fully trained classification model with a WOL of over 200k neurons and Softmax function. However, suppose we have an oracle that retrieves a set of neurons in a WOL with two properties: 1) The neuron representing the correct label is in the set and, 2) all other neurons in the retrieved set are likely to have a smaller inner product than the correct label. With this retrieval oracle in place, even if the correct label does not have the highest logit in the full prediction, it would be the highest one in this retrieved set, leading to even better accuracy than full softmax.

Therefore, based on the above observation, we design a superior retrieval mechanism, Label Sensitive Sampling (LSS), which enforces the two objectives mentioned above. The construction of this mechanism leverages the pre-trained model and its corresponding training dataset. We summarize our contributions as follows:
 \begin{itemize}[leftmargin=*,nosep,nolistsep]
    \item We identify an objective gap between efficient inference and the classical MIPS formulation. Moreover, to bridge this gap, we observe that inference over a perfect subset of output neurons is both efficient and accurate.
    
    
    
    \item Based on the above observation, we propose Label Sensitive Sampling (LSS), a hashing-based method that uses a learning mechanism to incorporate ground truth information in the retrieval function. This retrieval mechanism can sample a small subset from WOL neurons with a high probability of including label neurons. 
    
    \item We provide rigorous evaluations of our method on four large benchmark datasets using two different model architectures. We show that our method achieves up to $\bf{5} \times$ speed up and at most $\bf{87\%} $ energy reduction without any loss in accuracy compared to full computation inference.
    
\end{itemize}

\section{Related Work}
\label{related}

\subsection{Efficient Inference for Wide Output Layers}
Many approaches have been developed for efficient inference over WOLs. Most of these methods can be categorized as an approximate MIPS problem. \cite{zhang2018navigating}  proposed a graph-based method that maps the database vectors in a proximity graph~\cite{tan2019efficient,zhou2019mobius} and outperforms traditional PCA~\cite{bachrach2014speeding} or  SVD~\cite{shim2017svd} approaches in language modelling tasks. However, graph-based methods severe performance degradation's in parallel settings because of the difficulty in batching the greedy walks over the graph. Meanwhile, \cite{morozov2018non} also mentioned the potential risks in the asymmetric transformation in \cite{zhang2018navigating} . On the other hand, several MIPS solvers\cite{guo2016quantization,wu2017multiscale} have been proposed for inference over WOLs. However, these solvers trade plenty of computation for accuracy and are both energy and time consuming, even with full parallelism. We provide a detailed literature review in Appendix A.

\subsection{Hashing Algorithms for Large Scale Learning}\label{Hashing}
Hashing based data structures are widely applied in machine learning tasks at scale~\cite{chen2018lshff,spring2020mutual}. In formal terms, we consider $\mathcal{H}$ as a family of hash functions that maps $\mathbb{R}^{D}$ to some set $\mathcal{S}$. 

\begin{defn} [\bf LSH Family]\ A family $\mathcal{H}$ is called $(S_0,cS_0,p_1,p_2)$-sensitive if for any two points $x,y \in \mathbb{R}^{D}$ and $h$ chosen uniformly from $\mathcal{H}$ satisfies:
    \begin{itemize}[leftmargin=*,nosep,nolistsep]
    \vspace{-1mm}
        \item if $Sim(x,y)\ge S_0$ then ${Pr}(h(x) = h(y)) \ge p_1$
        \item if $ Sim(x,y)\le cS_0$ then ${Pr}(h(x) = h(y)) \le p_2$
    \end{itemize}
    \label{def:lsh}
\end{defn}

Here $Sim(x,y)$ is a similarity measure and $p_1 > p_2$ and $c < 1$ is required. Details are presented in Appendix A. The general idea of these LSH functions is to pre-partition the dataset into buckets where vectors within the same bucket are similar~\cite{shrivastava2014asymmetric,Proc:Indyk_STOC98,indyk2006polylogarithmic}. Therefore, given a query vector, the computation can be focused on a tiny subset of the large database. Taking advantages of this massive computation reduction, hashing methods have been applied in: (1) Feature representation: \cite{li2011hashing,li2012one} demonstrate an efficient way of performing efficient linear learning via permutation-based hashing that preserves Jaccard Similarity preserved (2) 
Neural network training: \cite{SLIDE} propose a sub-linear deep learning engine that use LSH to select neurons in forward and backward pass of NN training and achieve outperforming efficiency on CPU compared to a Tensorflow implementation on GPU. (3) Fast nearest neighbor search: \cite{shrivastava2014defense,wang2017flash} provide algorithms that tackle efficiency bottlenecks in metric similarity search on ultra high dimensional space.

\vspace{-2mm}

\begin{figure}[t]
    \centering
    \includegraphics[width= \textwidth,trim={1.5cm 10cm 3cm 6cm},clip]{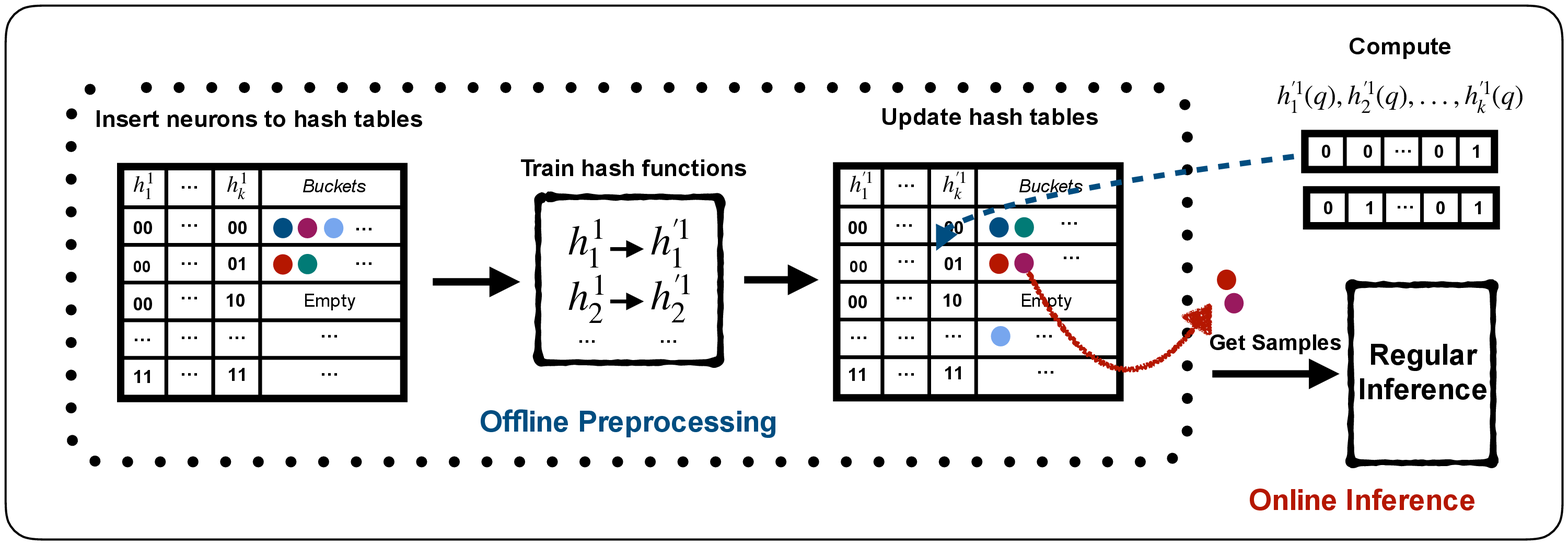}
    \caption{The LSS pipeline in two stages: 1) During preprocessing, we incorporate label information from training data in LSH hash functions and rebuild hash tables accordingly. 2) During the actual inference, the softmax computation of the WOL is based on a subset of neurons retrieved by the input embeddings from hash tables rather than the full set of neurons.}
    \label{fig:workflow}
\end{figure}
\vspace{-2mm}
\section{Bridging the Gap Between MIPS and NN Inference}

\subsection{Notation and Settings}
In the WOL setting, we denote the WOL's weight matrix as $W\in \mathbb{R}^{m \times d}$ and its bias vector as $b \in \mathbb{R}^m$, where $m$ is the size of output layer (number of classes) and $d$ is the embedding dimension. The WOL can be represented by a set of neurons $\mathcal{C} = \{c_i\,|\,0 \leq i < m \}$, where each neuron constitutes $w_i$, the $i^{th}$ row of $W$, and $b_i$ the $i^{th}$ element of $b$. Typically, $N\gg d$ for a wide output layer. During inference, given an input embedding $q \in \mathbb{R}^d$ from the previous hidden layer, the output of a forward pass through the WOL is $\sigma(qW^T+b)$, where $\sigma$ is some activation function that translates the logits into probabilities; then, the indices of the largest logits are returned as the predicted classes. This formulation of WOLs can be applied to the softmax output layer in language modeling and extreme classification, as well as the matrix factorization in collaborative filtering~\cite{xu2018deep}.


Our first goal is to construct the retrieval oracle introduced in Section~\ref{sec:intro}. We formulate the objective of the oracle as the construction of a \textbf{Perfect Retrieval Set}. For each input embedding (query), we sample a subset $\mathcal{S}$ of neurons, such that $k = |\mathcal{S}|$ and $k << N$. In the sampling process, we want to maximize the probability of retrieving label neurons. Moreover, label neurons should have the highest inner products within the subset. Formally,

\begin{defn}[Perfect Retrieval Set]\label{prob:lss}
Given a WOL with neurons $\mathcal{C}=\{c_i\,|\,i=1,2,...m\}$, for each input embedding $q$, with labels $y$ in the multi-label setting, we want to sample a subset $\mathcal{S} \subset \mathcal{C} $ with size $k \ll m$ such that,
$\argmax\limits_{c_j \in \mathcal{S}} q^{T} w_j \in y$. 
\end{defn}

\vspace{-2mm}
\subsection{Algorithm Overview}
To construct our retrieval oracle, which ideally returns a Perfect Retrieval Set, we introduce Label Sensitive Sampling (LSS). LSS is a scheme that exploits Locality Sensitive Hashing (LSH) along with an efficient hyperplane (hash function) training procedure to approximate the maximum inner product in WOLs using label information. Note that we choose a particular variant of LSH called SimHash~\cite{charikar2002similarity}, which is parameterized by hyperplanes in the dimensionality of the query. The full workflow is illustrated in Figure~\ref{fig:workflow}.

LSS works in two separate phases: an offline preprocessing stage (shown in algorithm~\ref{algo:pre}) and then an online inference stage (shown in algorithm~\ref{algo:inference}). We summarize the \textit{offline} construction as the following three steps: (1) Given the weight of a trained model, we build Locality Sensitive Hash Tables($HT$) via hash functions ($h$) constructed by (initially) randomly generated hyperplanes. (2) Based on the neurons retrieved from our initial hash tables by each query, we iteratively update the hyperplanes (hash functions) with our designed sampling loss. For the \textit{online} inference phase, we summarize LSS as the following two steps: (1) Given input from the test set, we compute the forward pass up to the last layer and get an embedding $q$. Then, we query the hash tables with $q$. (2) We set retrieved neurons as ``active'', and all other neurons as inactive for this input. Finally, we perform prediction on the ``active'' neurons and the top-ranked neurons (with highest logits) are returned as the prediction.

{\bf Why Hashing based Indexing:} In this work, we index the neurons (only Ids) to hash tables before performing LSS. There are three major advantages of using hash tables as the data structure for efficient inference: (1) \textit{Efficiency}: In the preprocessing phase introduced in Appendix A, hash tables have lower time and space complexity compared to tree or graph methods~\cite{ann}. In the query phase, hash table lookup operations are also faster than a greedy walk on tree or graph structures. (2) \textit{Differentiability}: The projection step in hash functions represents a space partition and can be adjusted via gradient-based methods~\cite{wang2017survey,hashnet}. (3) \textit{Scalability}: Compared to MIPS solvers~\cite{guo2016quantization} and graph methods~\cite{morozov2018non,zhang2018navigating}, hashing methods are more amenable towards less computation and multi-threading ~\cite{wang2017flash}. Therefore, hashing methods are capable of massive parallel inference on CPUs.



\vspace{-3mm}

\subsection{Preprocessing: Construction of Hash Tables }

Given a trained model, we construct $L$ hash tables, where each hash table has a capacity of $2^K$, and insert each WOL neuron into each hash table. We do so using $K$ binary hash functions per table (the hash table keys are constructed by concatenating the $K$ binary hashes together). In total, we need $K \times L$ hash functions. Specifically, each of the $K$ hash bits of an input $x$ is generated by function $f(x) = sign(\theta^Tx)$, where each column of $\theta_i \in \mathbb{R}^{d+1}$ is drawn i.i.d. from $\mathcal{N}(0,1)$. This is equivalent to the method used in Simhash~\cite{charikar2002similarity}. Geometrically, each $\theta_i$ ($i = 1,\dots,K \times L$) represents a projection hyperplane in $\mathcal{R}^{d+1}$, such that the space is partitioned by $K$ hyperplanes.

\begin{wrapfigure}{}{0.55\textwidth}
  \vspace{-0.3in}
    \begin{minipage}{0.55\textwidth}
\begin{algorithm}[H]
    \begin{algorithmic}[1]
        \STATE \textbf{Input: $Q$, $Y$, $W$, $HT$, $H$, $t_1$, $t_2$ }
        \STATE $P_+ = \{\}$, $P_- = \{\}$
        \FOR {$i=1:N$}
        \STATE Compute $h(q_i)$.
        \STATE S = $\{\}$
        \FOR {$l=1:L$}
        \STATE $S$ = $S \cup$ \textbf{Query}($h_l(q_i)$, $ht_l$)
        \ENDFOR
        \STATE $P_+ = P_+ \cup \{(q_i, w_{y_i}) | y_i \notin S , q_i^{T}w_{y_i}>t_1\}$
        \STATE $P_- = P_- \cup \{(q_i, w_j) | w_j \in S \setminus y_i , q_i^{T}w_j<t_2\}$
        \ENDFOR
        \STATE \textbf{shuffle} $P_+, P_-$
        \STATE $g = min(|P_+|, |P_-|)$
        \STATE $H^{'} \xleftarrow{} \textbf{IUL}(H, Pair_{p}[:g], Pair_{n}[:m])$
        \STATE $HT^{'} \xleftarrow{} \textbf{rebuild} (HT, H^{'})$
        \RETURN $HT^{'}$, $H^{'}$
    \end{algorithmic}
    \caption{Preprocessing}\label{algo:pre}
    \end{algorithm}
    \end{minipage}
    \vspace{-0.1in}
\end{wrapfigure}

During initialization, we insert neuron Ids into each hash table. Recall each neuron $c_i$ can be represented by the concatenation of its weight and bias parameters, $c_i = [w_i, b_i]$. Therefore, for each $[w_i, b_i]$, we generate $K \times L$ hash codes, which constitute $L$ total hash table keys, and insert $c_i$ into $L$ hash tables accordingly. For each input in the training set, we collect its embedding $q$ before it is fed forward to the WOL. Then, $[q, 0]$ serves as an input embedding query to retrieve corresponding neuron Ids from the hash tables. For simplicity, we omit $[w_i, b_i]$, and $[q, 0]$ and directly use $w_i$ and $q$ in the following sections.

In order to possess the properties of a Perfect Retrieval Set, the hash functions should have the following properties: (1) the collision probability between the input embedding query and its ground truth label neuron is high (2) the collision probability between the input embedding query and its non-label neurons is low (3) neurons are distributed evenly over all buckets for better load-balancing, which leads towards lower overhead (otherwise no efficiency gain). We formally define our ideal hash function as the following: 
\begin{defn} [\bf Label Sensitive Hash Family]\ A hash family $\mathcal{H}$ is called $(1,0,p1,p2)$-sensitive if for a triplet $(q,x,y) \in \left(\mathbb{R}^{d+1}\times \mathbb{R}^{d+1}\times \{0,1\}\right)$, a hash function $h$ chosen uniformly from $\mathcal{H}$ satisfies:
    \begin{itemize}[leftmargin=*,nosep,nolistsep]
    \vspace{-1mm}
        \item if $y=1$ then ${Pr}(h(q) = h(x)) \ge p_1$
        \item if $y=0$ then ${Pr}(h(q) = h(x)) \le p_2$ 
    \end{itemize}
    \label{def:lsh}
\end{defn}

We approximate hash functions from such a family based on an iterative learning mechanism, which encourages the above three properties using an Index Update Loss (IUL). The key to this learning process is the collection of positive and negative pairwise training samples. For each input embedding $q$, we retrieve its corresponding set of neurons $\mathcal{S}$ from the existing $L$ hash tables. Then, pairwise training samples $(q, w_i)$ are collected according to the following criterion:
\begin{itemize}[leftmargin=*,nosep,nolistsep]
\vspace{-1mm}
    \item positive pair $P_+ = (q, w_i) \text{ if } w_i \in C\setminus S \text{ and } w_i \in y \text{ and } q^Tw_i > t_1$
    \item negative pair $P_- = (q, w_i) \text{ if } w_i \in S \text{ and } w_i \notin y \text{ and } q^Tw_i < t_2$
\end{itemize}
{\bf Difference from Standard Learning to MIPS:} The positive and negative set construction is essential. It should be observed that standard learning approaches, focused on MIPS objective and use every positive and negative pair for training the hash function, potentially solving a harder problem. Instead, we only use the negative pairs arising from the buckets, and positive pairs missed by buckets if they are the correct labels. Overall, our training is aware of the retrieval mechanism and only enforces what is needed for classification.\\ 
{\bf Index Update Loss (IUL): } We use Hamming distance as an approximation of the difference between the hash codes of one training pair. Since $sign$ is a discrete function, we use $\tanh$ as a differentiable approximation. We know that $\text{dist}_{\text{hamming}}{ (q, w)} = \frac{1}{2}( K - q^Tw) $~\cite{hashnet}. Therefore, if the inner product between the query embedding $q$ and a particular neuron $w_i$ is high, they tend to have similar hash codes. Thus, we propose an Index Update Loss based on Hamming distance to update hash functions (random hyperplanes) with collected $P_+, P_-$. Formally,
\begin{equation}
   \mathcal{IUL}(P_+, P_-) = \sum_{q_i, w_i \in P_+}-log(\sigma( \mathcal{K}(w_i)^{T} \mathcal{K}({q_i}))) - \sum_{q_j, w_j \in P_-}log(1-\sigma( \mathcal{K}(w_j)^{T} \mathcal{K}(q_j))),  
\end{equation}
where $\mathcal{K}(w) = \tanh(\theta^Tw) $, $ \mathcal{K}(q) = \tanh (\theta^Tq) $, and $\sigma(x) = 1 / 1 + e^{- x}$.

 The intuition behind our IUL design is that positive pairs are encouraged to land in the same bucket while negative pairs are pushed towards different buckets. Positive samples are used to maximize the probability of including correct label neurons. Concurrently, it increases the relative ranking of label neurons on the inner product by decreasing the probability of retrieving other non-essential neurons. Negative samples are used to maintain a relatively small bucket size by pushing out low inner product neurons from the bucket. Otherwise, all the neurons would ultimately converge to the same bucket in each table. We collect the pairwise training data based on each retrieved set because it directly reflects the circumstances on how data are separated by the current hyperplanes. $t_1$ and $t_2$ are two inner product ranking thresholds, that control the inner product quality of positive and negative pairs. Usually, we have $t_1 > t_2$ in any valid setting. Otherwise, in the situation that a certain label neuron has a small logit, due to the nature of LSH, it would be challenging to train hyperplanes in the manner that low inner product neurons are retrieved while high inner product neurons are excluded. 

  \begin{wrapfigure}{}{0.4\textwidth}
  \vspace{-0.3in}
    \begin{minipage}{0.4\textwidth}
\begin{algorithm}[H]
    \begin{algorithmic}[1]
        \STATE \textbf{Input: $q$, $W$, $HT^{'}$, $H^{'}$ }
        \STATE Compute $H^{'}(q)$
        \STATE S = $\{\}$
        \FOR {$l=1:L$}
        \STATE $S$ = $ S \cup$ \textbf{Query}($H^{'}_l(q)$, $HT^{'}_t$)
        \ENDFOR
        \RETURN $qW_S^{T}$
    \end{algorithmic}
    \caption{Inference of one sample}\label{algo:inference}
\end{algorithm}
    \end{minipage}
    \vspace{-0.1in}
\end{wrapfigure}

\subsection{Online Efficient Inference}
After presenting the most essential component of our proposal, we introduce the online inference process. The model weights, hash functions, and hash tables are frozen before the inference. For each input in the testing set, the output embedding from the second to the last layer is first computed. The $K \times L$ hash codes of the embedding are generated to retrieve the corresponding neurons from the hash tables. The next step is similar to the usual inference routine of WOL for making predictions. However, instead of performing the full inference, the model only computes the logits of retrieved neurons.

\vspace{-2mm}
\section{Evaluation}


In this section, we evaluate the effectiveness of our proposed LSS method in efficient inference for WOL on two large scale extreme classification and two language modeling datasets. Specifically, we would like to answer the following questions: (1) Does LSS outperform other efficient inference approaches on energy and time? (2) How do the inner metric change during the learning process of LSS? (3) Can LSS always surpass the accuracy of full inference in a shorter time? 

\textbf{Datasets and Models:} For extreme classification, we use a standard fully connected neural network with one hidden layer of size 128. We evaluate on two datasets:  Wiki10-31K~\cite{wiki10} and Delicious200K~\cite{deli200k}. For language modeling, we use a standard fully connected network with one hidden layer of size 128 for the Text8~\cite{text8}, and a two-layer LSTM network with a hidden dimension size of 200 for wiki-text-2~\cite{wikitext2}. We present more experiment details in Appendix B.

\textbf{Baselines:} We compare the proposed LSS against the following state-of-the-art methods: (1) \textbf{SLIDE}~\cite{SLIDE} is a deep learning system utilizing locality-sensitive hashing for faster training, written in C++. We implement this method for inference. (2) \textbf{Graph Decoder} (GD) is a MIPS method proposed for efficient Softmax inference in~\cite{zhang2018navigating} that combines the asymmetric transform in ~\cite{bachrach2014speeding} with HNSW~\cite{malkov2018efficient}. Here we exploit the original implementation of HNSW~\cite{HNSW} and pre-process the data according to ~\cite{zhang2018navigating}. (3) \textbf{ip-NSW} is a state-of-the-art graph-based MIPS algorithm proposed in~\cite{morozov2018non, ipnsw}. It belongs to the direct MIPS category and shows performance improvement over GD. (4) \textbf{Product Quantization }(PQ)~\cite{pq} is a MIPS solver with K-means and asymmetric transformation. We implement this method following the popular open-source ANNS platform from Facebook \cite{JDH17}. (5) \textbf{FULL} is the regular but paralleled NN inference using all neurons in the last layer.

\textbf{Implementation and Experiment Setting:} All the experiments are conducted on a machine equipped with two 20-core/40-thread processors (Intel Xeon(R) E5-2698 v4 2.20GHz). The machine is installed with Ubuntu 16.04.5 LTS. LSS for the output layer is written in C++ and compiled under GCC7 with OpenMP. The full inference is implemented in PyTorch. GD, ip-NSW, PQ are implemented in C++ with OpenMP. All implementation is parallelized with multi-threading with full usage of CPU cores. All baselines use the best results after extensive hyperparameter search. CPU energy consumption is monitored over time with the command line tool described in Appendix C. 

\textbf{Evaluation Metric:}
We compare our method against other baselines from multiple evaluation metrics: (1) \textbf{Precision@k} (P@k) for multi-label classification tasks. (2) \textbf{Label Recall} indicates the proportion of the correct labels in the retrieved ones. (3) \textbf{Time} is measured as the average wall-clock time for passing 1000 testing data through the last layer in seconds. (4) \textbf{Energy Consumption}, measured in Joules, is the average CPU power (Watts) over the inference period, multiplied by the inference time. It is then averaged for every 1000 samples. (5) \textbf{Collision Probability} indicates the probability that a pair of inputs are hashed to the same bucket for a fixed hash table. We expect positive pairs to have high collision probabilities and negative pairs to have low collision probabilities.

\vspace{-2mm}
\subsection{Main Result}\label{main:res}
In this section, we compare LSS with baseline methods for the trade-off between accuracy and efficiency. For each method, we aim to minimize the time and energy spent in the inference while maximizing the $P@1$ and $P@5$. Following this strategy, we report the best performance of LSS and all other baselines on four datasets in Tables  \ref{result:deli200K}, \ref{result:text8}, \ref{result:wiki10}, \ref{result:wikitext2}. From these tables, we observe that: (1) LSS achieves the best $P@1$ and $P@5$ compared to other methods.  (2) The sample size of the LSS method is the smallest. LSS uses at most 6\% of the neurons in the output layer. Furthermore, we can see that on a larger dataset, LSS samples even fewer neurons. In Delicious200K, the output space is over $200K$, while LSS only uses 360 neurons for inference computation. In Text8, the output dimension is 1,355,336, while LSS only uses 965 neurons on an average. LSS can match full accuracy with much less computation. (3) Most importantly, we observe that LSS achieves up to $5.1 \times $ reduction in time and $8.2 \times $ reduction in energy consumption. 

The experimental results validate our argument regarding the gap between MIPS and inference, as well as our choice of using hash tables. (1) We observe that MIPS approximation algorithms usually have a low label retrieval rate. (2) Even though other baselines only visit a small portion of neurons, they fail to achieve consistent speed up. These observations validate our reason for choosing a hashing-based approach, as it is easier to parallelize. Previous works~\cite{zhang2018navigating,chen2018learning} compared the performances of different methods under a single CPU thread setting, which is not a practical simulation for real-world cloud systems. Moreover, methods such as ip-NSW or GD are ill-suited to exploit the full parallelization offered by multi-core CPUs and tend to have a large number of irregular memory accesses. This limitation significantly degrades their performance even compared to exact MIPS computation on 
CPU with the current PyTorch framework. 

Based on the above results, we answer the first question from the beginning of the section: compared to full inference, LSS can perform inference with comparable accuracy using only 12$\%$ energy and $19\%$ time. Moreover, even in scenarios where full inference is extremely parallelizable and outperforms all other approximate MIPS approaches, LSS still achieves the best efficiency.

\begin{table*}[t!]
\caption{Baseline comparisons on various datasets}
\label{table:main}
\LARGE
\centering
\captionsetup[sub]{labelfont=small,textfont=small}
\begin{subtable}[t]{0.57\textwidth}
\subcaption{Results for Delicious200K}
\centering
\resizebox{\linewidth}{!}{
\renewcommand{\arraystretch}{1.23}
\begin{tabular}{|c||c|c|c|c|c|c|}
        \hline
         & LSS & Full & PQ & ip-NSW & GD \\\hline\hline
        p@1 & 0.4245 & 0.4391 & 0.1079 & 0.0693 & 0.4362\\\hline
        p@5 &  0.3473 & 0.3619 &0.1180 & 0.0256 & 0.3581\\\hline
        Sample size &  424 & full & full & 3000 & 3000\\\hline
        Label Retrieval Rate &  0.889 & 1 & 0.3464 & 0.0900 & 0.7000\\\hline \hline
        \multirow{1}{*}{\makecell{Avg. Time \\ Per 1000 samples(s)}} &  \bf{0.81(5.1x)}&    4.16&    10.51&    2.45&    2.29
        \\[16pt]\hline
         \multirow{1}{*}{\makecell{Avg. Energy \\ Per 1000 samples (J)}} & \bf{8.70 (8.2x)}&    71.34&    116.61    &33.88&    29.05
         \\[16pt]\hline
\end{tabular}
}
\label{result:deli200K}
\end{subtable}
\begin{subtable}[t]{0.42\textwidth}
\subcaption{Results for Text8}
\resizebox{0.98\linewidth}{!}{
\renewcommand{\arraystretch}{1.15}
\begin{tabular}{|c|c|c|c|c|c|}
        \hline
        LSS & Full & PQ & ip-NSW & GD \\\hline\hline
        0.9132 &  0.9129 & 0.1631 & 0.8299 & 0.9129\\\hline
        0.7404 & 0.7370 & 0.1631 & 0.6652 & 0.7370 \\\hline
         965 & full & full & 3000 & 3000\\\hline
        1 & 1 & 0.5842 & 0.8977 & 0.9908\\\hline \hline
        \multirow{1}{*}{\makecell{\bf{0.56(3.3x)} \\}}    &1.88&    13.92&    2.07&    2.09
        \\[16pt]\hline
        \multirow{1}{*}{\makecell{\bf{4.98(6.4x)}\\}}&    31.99    &174.49&    22.66&    20.40\\[16pt]\hline
\end{tabular}
}
\label{result:text8}
\end{subtable}

\bigskip

\begin{subtable}[t]{0.57\textwidth}
\subcaption{Results for Wiki10-30K}
\centering
\resizebox{\linewidth}{!}{
\renewcommand{\arraystretch}{1.23}
\begin{tabular}{|c||c|c|c|c|c|}
        \hline
          & LSS & Full & PQ & ip-NSW & GD \\\hline\hline
        p@1   & 0.8018 & 0.8232 & 0.3309 & 0.3207 & 0.7636\\\hline
        p@5  & 0.4822 & 0.5700 & 0.3259 & 0.1603& 0.4790 \\\hline
        Sample size & 559 & full & full & 1500& 3000 \\\hline
        Label Retrieval Rate  & 0.9779 & 1 & 0.8905 & 0.4854 & 0.9163 \\\hline \hline
        \multirow{1}{*}{\makecell{Avg. Time \\ Per 1000 samples(s)}}  & \bf{0.39(1.9x)}&    0.76    &4.06&    1.65&    1.69\\[16pt]\hline
        \multirow{1}{*}{\makecell{Avg. Energy \\ Per 1000 samples (J)}}  & \bf{3.53(3.0x)}&    10.69&    39.28&    15.75&    15.80\\[16pt]\hline
\end{tabular}
}
\label{result:wiki10}
\end{subtable}
\begin{subtable}[t]{0.42\textwidth}
\subcaption{Results for Wiki-Text2}
\centering
\resizebox{0.98\linewidth}{!}{
\renewcommand{\arraystretch}{1.15}
\begin{tabular}{|c|c|c|c|c|}
        \hline
        LSS & Full & PQ & ip-NSW & GD \\\hline\hline
        0.4265  & 0.4044 & 0.2234 & 0.0750 & 0.1369 \\\hline
        0.0837& 0.0774 & 0.0430 & 0.0271 & 0.0478 \\\hline
        3071 & full & full  & 5365  & 4956 \\\hline
        0.9284 & 1 & 0.6654  & 0.8705 & 0.9215 \\\hline \hline
        \multirow{1}{*}{\makecell{\bf{0.36(1.7x)}\\}}&    0.63&    10.57&    1.60&    1.76
        \\[16pt]\hline
       \multirow{1}{*}{\makecell{\bf{3.20(2.9x)}\\}}&    9.31&    128.76&    23.92&    27.07
        \\[16pt]\hline
\end{tabular}
}
\label{result:wikitext2}
\end{subtable}
\end{table*}
\begin{figure*}[t!]
    \captionsetup{labelfont=normalsize,textfont=normalsize}
    \begin{center}
        \begin{tabular}{ll}
            \includegraphics[width=0.48 \textwidth]{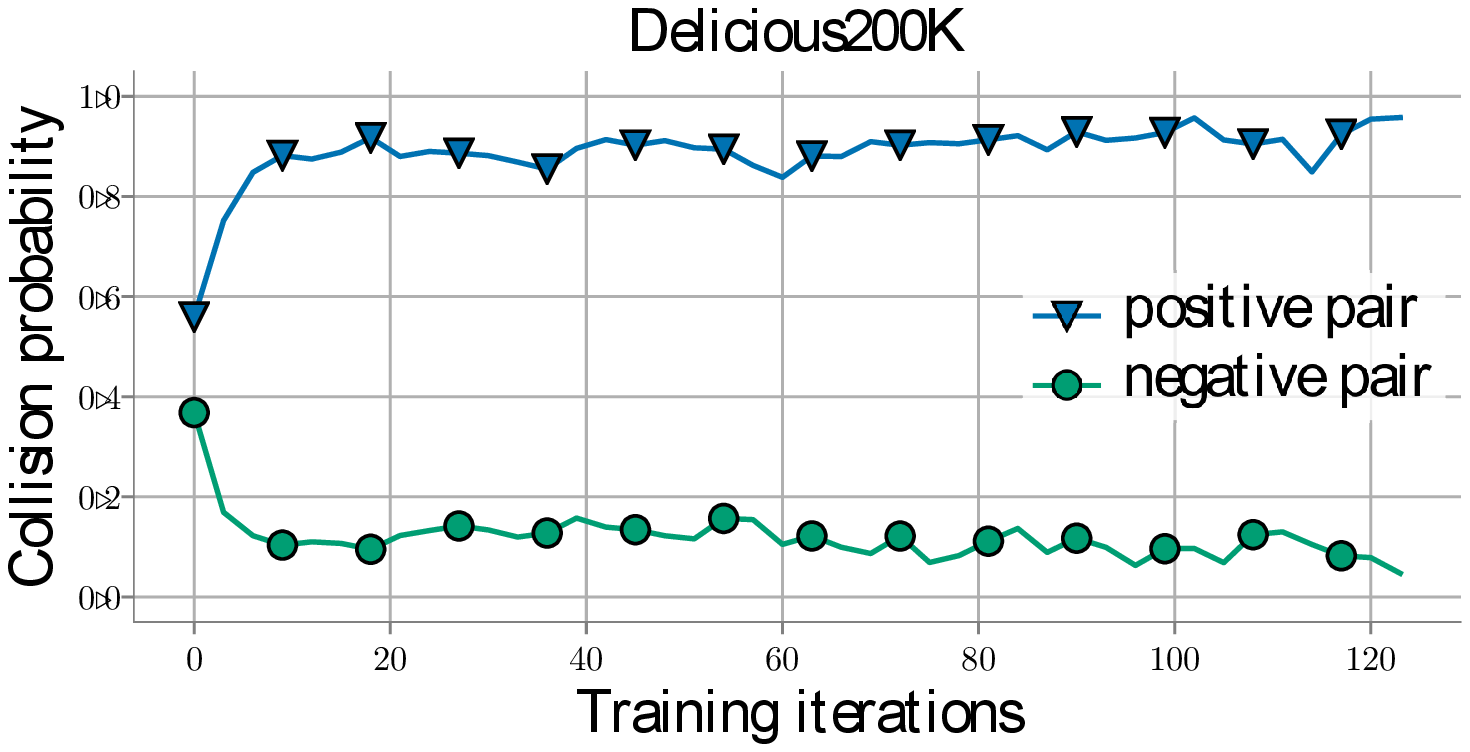} 
            \includegraphics[width=0.48 \textwidth]{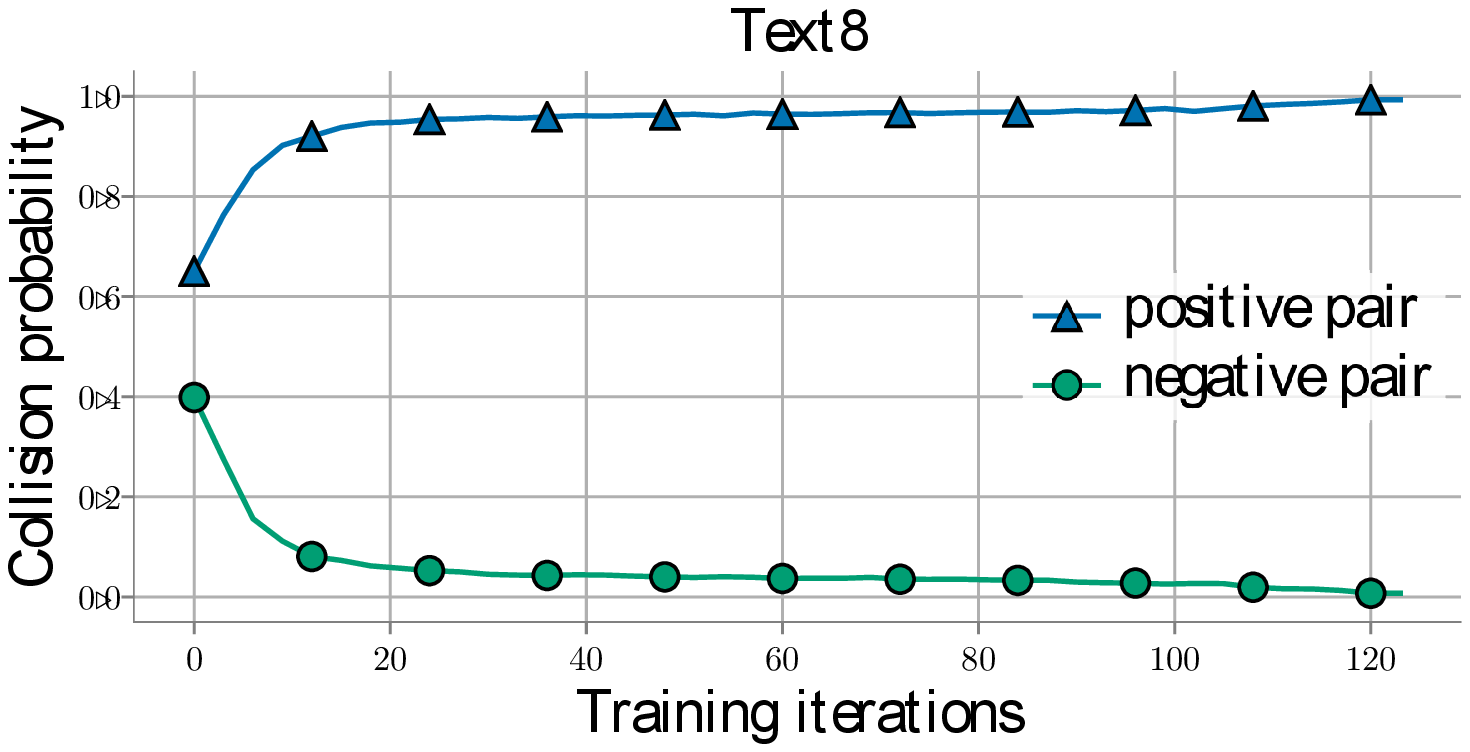} \\
        \end{tabular}
    \end{center}
     \vspace{-4mm}
    \caption{Blue line plots the collision probability between positive pairs. Green line plots the collision probability between negative pairs. For all the experiment, hashing function training batch is 256. The starting point represents the probabilities of random simhash.}\label{fig:freq}
    \vspace{-4mm}
    \label{fig:cp} 
\end{figure*}

\vspace{-2mm}
\subsection{Inner Metrics of LSS}
\textbf{Collision Probability} We investigate the process of training LSS hash functions. Figure \ref{fig:cp} shows the collision probability for both positive and negative pairs we collect. We observe that, in both Text8 and Delicious200K, collision probability between positive pairs increases and converges to a level above 0.9. Meanwhile, the collision probability between negative pairs decreases throughout the training process. In Delicious200K, the collision probability of negative pairs converges to around 0.1 while in Text8, it converges close to zero. This observation helps to explain the significantly higher label retrieval rate of LSS: the learned hash functions identify projections that assign higher collision probabilities of label neurons with a high inner product. As we mentioned in section~\ref{sec:intro}, the average inner product rank of label neurons is only 498/205443. With the LSS learned hash functions, we achieve a significantly higher label neuron inner product rank of 7.77. 

\textbf{Choosing K and L:} We investigate the choice of $K$, the number of projections, and $L$, the number of hash tables, on Delicious200K. $K$ and $L$ are the two main hyperparameters affecting computation time and accuracy. The objective of this experiment is to establish the robustness of LSS's accuracy with various sample sizes. This robustness directly relates to the trade-off between accuracy and inference efficiency. As reported in table \ref{tab:klabla}, $L = 1$ and $K = 4$ leads to the most efficient inference with tolerable accuracy loss. This hyperparameter is chosen because it requires less hash code computations (determined by $K$), a fewer number of table lookups (determined by $L$), and smaller last layer matrix multiplication (determined by sample size). On the other hand, we observe that $P@1$ and $P@5$ do not vary too much with modification of $K$ and $L$. This experimental phenomenon suggests that the learning objectives can consistently guide the LSS towards a set of retrieved neurons achieving decent $P@1$ and $P@5$, independent of the parameters of hash table structures.

\begin{table*}[t]
\vspace{-0.1in}
\caption{Effect of $L, K $ on $P@1/5$ for Delicious-200K dataset.}
\resizebox{\linewidth}{!}{
\renewcommand{\arraystretch}{1.15}
\begin{tabular}{|c|ccc|ccc|ccc|}
\hline
 & \multicolumn{3}{c|}{K=4} & \multicolumn{3}{c|}{K=6} & \multicolumn{3}{c|}{K=8}\\
\cline{2-10} &P@1&P@5 & Sample Size  & P@1 & P@5 & Sample Size & P@1 & P@5 & Sample Size  \\
\hline
L=1 & 0.4245 & 0.3473 & 424  & NA & NA & 0    & NA & NA & 0  \\
\hline
L=10 & 0.4602 & 0.3676  & 2560 & 0.4488 & 0.3733 & 875.53 & 0.4408 & 0.3598& 153.31\\
\hline 
L=50 & 0.4405 & 0.3659 & 15568 & 0.4455  & 0.3599  & 2122.47 & 0.4457 & 0.3615 & 360 \\
\hline
\end{tabular}
}
\label{tab:klabla}
\vspace{-0.3cm}
\end{table*}

\textbf{Energy Efficiency:} We summarize energy usages of LSS in table~\ref{table:main}. On all datasets, we observe significant energy reduction. To further distinguish the learning mechanism from the naive data structure, we measure the energy usage of SLIDE, which exploits random Simhash. Given the Wiki10-30K dataset, we initialize the hash tables with random projection and then vary $K$ and $L$ for performance. Using the same optimization criterion described in Section ~\ref{main:res}, it retrieves 741 neurons, and on average, it takes 0.68 ms and 9.89 J to process 1000 samples. Compared to full inference, SLIDE requires less time and energy. This reduction demonstrates the benefit of hashing based retrieval functions. Hashcode calculation and hash table lookup are both energy-efficient operations. On the other hand, LSS achieves better performance compared to SLIDE. This reduction directly relates to smaller sample size and demonstrates the power of our learning mechanism.


Based on the analysis of internal metrics, we answer the second question: The learning process in LSS generates better projections that benefit for: (1) Label neurons retrieval, (2) Robustness over hash table parameters, (3) Energy and time improvements.

\subsection{Accuracy Advantage}
\vspace{-0.1in}
We demonstrate the potential of surpassing full inference accuracy with the right retrieval mechanism. Here we modify the optimization direction. We aim to find LSS parameters that outperform full inference in accuracy with sub-sampled neurons instead of finding the optimal accuracy-efficiency trade-offs. Table \ref{tab:highacc} presents the highest accuracy LSS achieves on each dataset. We notice that on Delicious-200K and Wiki-Text2, LSS can outperform full computation accuracy. For Wiki10-31k and Text8, we achieve matching $P@1$ and $P@5$. Based on these results, we answer the third question: LSS is capable of surpassing full inference via sampling an subset of neurons for each input embedding. It suggests LSS's strength in distinguishing label neurons.

\begin{table}[hbt!]
\vspace{-0.1in}
    \centering
    \caption{This table summarizes the highest accuracy of LSS on four dataset}
    \resizebox{0.7\linewidth}{!}{
    \centering
    \begin{tabular}{|c||c|c|c|c|}
        \hline
        Dataset  & Wiki10-31k & Delicious200k & Text8 & wiki2 \\\hline
        LSS p@1   & 0.8232 & \bf{0.4596} & 0.9129 & \bf{0.4265} \\\hline
        LSS p@5  & 0.4822 & \bf{0.3676} & 0.7370 & \bf{0.0837} \\\hline
        sample Size & 2372 & 1487 & 1008 & 3071  \\ \hline  \hline
        Full p@1 & 0.8232 & 0.4391 & 0.9129 & 0.4044 \\\hline
        Full p@1  & 0.5700 & 0.3619 & 0.7370 & 0.0774  \\\hline 
        
    \end{tabular}
    }
    \label{tab:highacc}
 \vspace{-0.2in}   
\end{table}


\section{Conclusion}
In this paper, we introduce LSS—a hashing based approach
that performs energy and time efficient inference on a wide output layered neural networks. We show a novel problem formulation that identifies and bridges the gap between efficient inference and maximum inner product search (MIPS).  We propose a combination of hashing based data structures and hyperplane learning objectives for efficient retrieval of label neurons. We present a novel index update loss that dynamically adapts the hash functions that reduce the sample size while preserving the prediction accuracy. We compare LSS against graph-based MIPS methods, direct MIPS solvers, and exact full inference on four real-world scenarios. We show that LSS substantially outperforms other MIPS baselines from both accuracy and efficiency metrics, with up to 8x energy reduction and up to 5x speedup compared to ideally parallelized full inference. 

\section*{Broader Impact}
This paper proposes a novel method for efficient inference on large output layer models, which have a wide range of applications in real-world settings such as recommendation systems and language models. According to Facebook, its deep neural network-based recommendation systems consume more than 70 \% of the cloud server's workload. It is more than evident that these large models require massive computation and lead to high electricity usage and CO2 emissions. We show that our method achieves 60-80 \% energy reduction with negligible accuracy loss. Our method is even orthogonal to the quantization direction, which has been recognized as the to-go way to reduce 30-40 \% energy usage. We believe our work takes a  solid step towards a more environmentally friendly and financially friendly machine learning.

\bibliographystyle{unsrt}
\bibliography{reference}
\newpage
\appendix
\section{Related Literature}
\label{related}

\subsection{Maximum Inner Product Search}

The application of deep neural network (NN) models in cloud services is usually associated with a Wide Output Layer (WOL). For recommendation NN models~\cite{Bhatia16,xue2017deep}, the size of the WOL is equal to the number of items to be recommended. For language modelling, the number of neurons in the WOL is equal to the vocabulary size. This large output space becomes the computation bottleneck and for this paper, we specifically focus on inference efficiency, which is a major concern for deployment in a cloud computing setting 

To tackle this inefficiency, various methods focus on efficient retrieval of the Top-K logits generated by the NN model. Most of these methods can be categorized as an approximate \emph{Maximum Inner Product Search} (MIPS) problem.  Formally, we aim to solve the following problem: given a set $S$ containing all neurons in the WOL as high dimensional vectors and each input embedding to the output layer as query $q$, we aim to develop an efficient algorithm for computing

\begin{equation}\label{eq:mips}
	w=\arg\max_{x\in S} x^\top q.
\end{equation}

There are two main categories for efficient inference         via MIPS. The first branch of methods aims to reduce the MIPS to classical approximate nearest neighbor search (ANNS) method~\cite{shrivastava2014asymmetric,shrivastava2014improved}, which can be summarized as two steps: (1) pre-processing the data vector $x\in S$ to $x^*$ and query vector $q$ to $q^*$ asymmetrically so that $x^Tq \approx f(x^*,q^*)$. Here $f(x,y)$ is cosine distance or euclidean distance. (2) perform ANNS via indexing structures such as quantization~\cite{JDH17}, or small world graph~\cite{malkov2012scalable,malkov2014approximate,malkov2018efficient}.  Another category of MIPS-based methods target at directly performing inner product search without reduction. This direct inner product search can be performed via graph due to the flexibility of modifying the edge definitions \cite{morozov2018non,zhou2019mobius,tan2019efficient}. 



\subsection{Locality Sensitive Hashing}\label{Hashing}

In this section, we briefly describe the recent development of using locality sensitive hashing~\cite{Proc:Indyk_STOC98,indyk2006polylogarithmic}. The high-level idea of LSH is to place similar items into the same bucket of a hash table with high probability. In formal terms, we consider $\mathcal{H}$ as a family of hash functions that maps $\mathbb{R}^{D}$ to some set $\mathcal{S}$. 

\begin{defn} [\bf LSH Family]\ A family $\mathcal{H}$ is called\\
    $(S_0,cS_0,p_1,p_2)$-sensitive if for any two points $x,y \in \mathbb{R}^{D}$ and $h$ chosen uniformly from $\mathcal{H}$ satisfies:
    \begin{itemize}[leftmargin=*,nosep,nolistsep]
    \vspace{-1mm}
        \item if $Sim(x,y)\ge S_0$ then ${Pr}(h(x) = h(y)) \ge p_1$
        \item if $ Sim(x,y)\le cS_0$ then ${Pr}(h(x) = h(y)) \le p_2$
    \end{itemize}
    \label{def:lsh}
\end{defn}

Typically, $p_1 > p_2$ and $c < 1$ is needed. Moreover, the algorithm uses two parameters, $(K, L)$. We construct $L$ independent hash tables from the collection. Each hash table has a meta-hash function $H$ that is formed by concatenating $K$ random independent hash functions. Given a query, we collect one bucket from each hash table and return the union of $L$ buckets. Intuitively, the meta-hash function makes the buckets sparse and reduces the number of false positives, because only valid nearest-neighbor items are likely to match all $K$ hash values for a given query. The union of the $L$ buckets decreases the number of false negatives by increasing the number of potential buckets that could hold valid nearest-neighbor items. One sufficient condition for a hash family $\mathcal{H}$ to be a LSH family is that the \emph{collision probability} ${Pr}_\mathcal{H}(h(x) = h(y))$ is a monotonically increasing function of the similarity, i.e. 
\begin{equation}\label{eq:monotonic}
\small
Pr_\mathcal{H}(h(x) = h(y)) = f(Sim(x,y)),
\end{equation} 
where $f$ is a monotonically increasing function.

The overall generation algorithm of nearest neighbor candidates works in two phases (See \cite{spring2017new,chen2018lsh} for details): 

{\bf Pre-processing Phase:} Constructing $L$ hash tables from the data by storing all elements $x \in \mathcal{C}$. We only store pointers to the vectors in the hash tables because storing whole data vectors is very memory inefficient. 

{\bf Query Phase:} Given a query $Q$, we search for its nearest neighbors. We obtain the union from all buckets collected from the $L$ hash tables. Note that, we do not scan all the elements in $\mathcal{C}$, we only probe $L$ different buckets, one bucket for each hash table. After generating the set of potential candidates, we compute the distance between the query and each item in the candidate set, and sort to find the nearest neighbor.


\section{Experiment Details}
\subsection{Dataset Statistic}\label{sec:datastats}

In our work, we present experiment on 4 datasets. 
The first two datasets, Wiki10-31k and Delicious-200K  are obtained from the Extreme Classification Repository\cite{Bhatia16}, which is a benchmark for various recommendation systems. Each Extreme Classification dataset uses  Bag-of-words (BoW) features as input and multi-hot label vector as output. For language modelling, we introduce 2 datasets from two models. For Word2vec model, we use the text8 dataset from~\cite{text8}. We conduct three preprocessing steps on the dataset: (1) Remove the words with frequency less than 2 from the vocabulary and mark the removed word as 'UNK'. (2) Represent each word in the document as input one-hot vector. (3) For each input word, represent its previous 25 words and after 25 words as a multi-hot vector. Then, use the vector as label. We also introduce a RNN based language models that uses the Wiki-Text-2 dataset~\cite{wikitext2}. In this dataset, we given a 35 word sequence as a multi-hot vector input, we would like to predict the next 35 words sequence. Therefore, the label vector is also multi-hot. Details about the datasets are shown in table below.

\begin{table}[h]\label{dataset}
\begin{center}
 \caption{\footnotesize Summary of output dimension for our benchmark dataset}
 \begin{tabular}{|c | c c c c c|} 
 \hline
 Dataset & Wiki10-31k & Delicious-200K & Text8 & Wiki-Text-2 &\\ [0.5ex] 
 \hline
 Output Dimension & 30938 & 205443 & 1355336 & 50000 & \\ 
 \hline
 Input Dimension & 101938 & 782585 & 1355336 & 50000 & \\ 
 \hline
 Training Samples & 14146 & 6616 & 11903644 & 725434 & \\ 
 \hline
 Testing Samples & 196606 & 100095 & 5101563 & 245550 & \\ 
 \hline
 
\end{tabular}
\end{center}
\end{table}
\subsection{Task and Models}\label{sec:model}

\textbf{Extreme Classifications}
The Extreme Classifications model targets at predicting the labels with ultra-high label Dimensionality given the input BoW features in ultra-high Dimensionality. The network architecture is summarized as: (1) Embedding layer that maps multi-hot input vector into a dense 128 dimension vector. (2) Relu actiation function. (3) Output layer with number of neurons equal to label Dimensionality.

\textbf{Word2vec}
The Word2vec model targets at predicting the neighbor words given the central word. The network architecture is summarized as: (1) Embedding layer that maps one-hot input vector into a dense 128 dimension vector. (2) Relu actiation function. (3) Output layer with number of neurons equal to vocabulary size.  

\textbf{RNN}
The RNN language model targets at predicting the next sequence of words given the current sequence of words. The network architecture is summarized from input to output as: (1) Embedding layer that maps multi-hot input vector into a dense 200 dimension vector. (2) First Dropout function. (3) 2 LSTM layers with hidden size equivalent to 200. (4) Second Dropout function.  (3) Output layer with number of neurons equal to vocabulary size. 

\subsection{IUL loss}\label{new:loss}
\begin{itemize}[leftmargin=*,nosep,nolistsep]
\vspace{-1mm}
    \item positive pair $P_+ = (q, w_i) \text{ if } w_i \in C\setminus S \text{ and } w_i \in y \text{ and } q^Tw_i > t_1$
    \item negative pair $P_- = (q, w_i) \text{ if } w_i \in S \text{ and } w_i \notin y \text{ and } q^Tw_i < t_2$
\end{itemize}
We use Hamming distance as an approximation of the difference between the hash codes of one training pair. Since $sign$ is a discrete function, we use $\tanh$ as a differentiable approximation. We know that $\text{dist}_{\text{hamming}}{ (q, w)} = \frac{1}{2}( K - q^Tw) $~\cite{hashnet}. Therefore, if the inner product between the query embedding $q$ and a particular neuron $w_i$ is high, they tend to have similar hash codes. Thus, we propose an Index Update Loss based on Hamming distance to update hash functions (random hyperplanes) with collected $P_+, P_-$. Formally,
\begin{equation}
   \mathcal{IUL}(P_+, P_-) =  - \sum_{q_i, w_i \in P_+}log(\sigma( \mathcal{K}(w_i)^{T} \mathcal{K}({q_i}))) - \sum_{q_j, w_j \in P_-}log(1-\sigma( \mathcal{K}(w_j)^{T} \mathcal{K}(q_j))), 
\end{equation}
where $\mathcal{K}(w) = \tanh(\theta^Tw) $, $ \mathcal{K}(q) = \tanh (\theta^Tq) $, and $\sigma(x) = 1 / 1 + e^{- x}$.
\section{Energy Measurement}\label{energy}
In our work, we measure the energy of inference methods via a monitoring tool. Command line utility tools, including \textit{s-tui}, were used to monitor the CPU power consumption, in Watts (Joules / second), over the inference times for each dataset and each method, in intervals of 1 second. The base power consumption would be subtracted from the average power of the inference time, in order to measure and compare the energy expenditure of only the inference step of each method.

\end{document}